\documentstyle[12pt]{article}

\begin{document}
\title{Relativistic Density Dependent Hartree-Fock
Approach for Finite Nuclei }
\author{Shi, Hua-lin and
Chen, Bao-qiu\\
China Institute of Atomic Energy, Beijing 102413, P.R.China\\
Ma , Zhong-yu \\
CCAST, P.O.Box 8730, Beijing 100080, P.R.China\\
China Institute of Atomic Energy, Beijing 102413, P.R.China}
\date{ }
\maketitle
\vspace{.5cm}
\begin{abstract}
  The self-energy of the Dirac Brueckner-Hartree-Fock calculation in
nuclear matter is parametrized by introducing density-dependent coupling
constants of isoscalar mesons in the relativistic Hartree-Fock (RHF)
approach where isoscalar meson $\sigma$ , $\omega$ and isovector meson $\pi$ ,
$\rho$
contributions are included. The RHF calculations with density
dependent coupling constants obtained in this way
not only reproduce the nuclear matter saturation properties , but also
provide the self-energy with an appropriate density dependence. The
relativistic density
dependent Hartree-Fock (RDHF) approach contains the features of the
relativistic G matrix
and in the meantime simplifies the calculation. The ground state properties of
spherical nuclei calculated in the RDHF are in good agreement with the
experimental data.
The contribution of isovector mesons $\pi$ and $\rho$ , especially the
contribution of the tensor coupling of $\rho$ meson , are discussed in
this paper.
 \\

\end{abstract}

\newpage
     \begin{center}
     \section{Introduction}
     \end{center}
     \begin{sloppypar}

   It is well known that the relativistic mean field (RMF) theory has been
extensively used to investigate the ground state properties of spherical and
deformed nuclei\cite{ser86,gam90} in the past several years. Recently , it is
used
to study the properties of nuclei far from $\beta$ stability
line\cite{tok91,sha94}.
Bouyssy et al\cite{bou87} extended the RMF to the relativistic
Hartree-Fock(RHF) approach.
The contributions of the exchange terms and isovector mesons in RHF to the
ground state properties of nuclei have been emphasized. Since then, the RHF
has been developed by several authors\cite{ber93,zha91}.

     \end{sloppypar}
     \begin{sloppypar}

   Though the RMF or RHF approach has been quite successful in reproducing the
bulk
properties of nuclei. However, too large compressibility of
nuclear matter is produced in these approaches . It might indicate an incorrect
density and
momentum behavior of the effective interaction described in the RMF and RHF
approaches. Some discrepancies in reproducing the properties of finite nuclei
in the standard RMF or RHF calculations are observed.
The calculations yield correct binding energies but too little charge radius
or vice versa. Such calculations reveal a new "Coester" band in a dependence of
1/$r_{ch}$
on $E_{B}$ which is similar as in the nonrelativistic Brueckner-Hartree-Fock
approach (BHF)\cite{kum78}. A more fundamental and sophisticated way to deal
with
the many-body problem is the Dirac Brueckner-Hartree-Fock(DBHF)
approach[9-12].
Starting from a bare nucleon-nucleon interaction of one boson exchange
potential,
one solves the Brueckner-Goldstone equation in the nuclear medium.
The Brueckner G matrix is the two nucleon effective interaction including
nucleon-nucleon short range correlation.
The DBHF has been quite successful in reproducing the nuclear matter saturation
properties, as well as the
compressibility\cite{haa87,bro84}. It provides a correct
description of the density- and momentum- dependence of the nucleon
self-energy in the nuclear medium. However, the DBHF due to its
complexity is mainly restricted to nuclear matter and only very few
finite-nuclei calculations  are performed so far\cite{mut90}.
Therefore, it is requested to remedy the deficiencies of the RMF or RHF
approach
with an effective interaction without losing the features of the relativistic
G matrix and at same time retain the simplicity.

     \end{sloppypar}

     \begin{sloppypar}

     Recently, there is a growing effort on developing the  relativistic
 effective interaction both in nuclear matter and  finite  nuclei.   Attempts
have
 been also made to improve both binding energy and rms  radii  of  finite
 nuclei  simultaneously  in the  RMF  or  RHF  with various effective
 interactions. Gmuca \cite{gmu91}  has  parametrized  the    results  of  the
DBHF
 in nuclear matter in terms of the RMF with  nonlinear  scalar
 and vector self-interactions. Though the coupling constants obtained in such
way
 is density-independent , these mesonic self-interactions implicitly represent
 the density-dependence.
 Brockmann and Toki \cite{bro92} developed a  relativistic
 density dependent Hartree (RDH) approach  for  finite  nuclei. The  coupling
constants
 of isoscalar meson $\sigma$ and $\omega$ in the RMF are adjusted at each
 density  by reproducing the nucleon self-energies  in nuclear matter
 resulting from the DBHF instead of fitting them to the empirical nuclear
 matter saturation properties. The binding energies  and  rms radii of
 $^{16}$O and $^{40}$Ca  calculated with the density dependent interaction
 are in good agreement with experiments. However,   Fritz, M\"{u}ther
 and Machleidt \cite{fri93} pointed out that the Fock  terms are not
 negligible. The relativistic density dependent Hartree-Fork(RDHF) calculations
 were performed \cite{fri93}. Due to the uncertainty of the DBHF at very
 low densities , the extrapolation procedure has to be adopted when
 applied to the calculation of finite nuclei.
 The sensitivity of the results calculated by the RDH or RDHF
 approaches on the extrapolation of the  coupling  constants  at  very  low
 densities are observed \cite{bro92,fri93}. However,  no  detail   procedures
of
 the extrapolation were presented there . Since there is  no  direct
restriction  on  the
 coupling constants at very low density , in Ref.\cite{ma94c} the scalar  and
 vector potentials of the DBHF  results were extrapolated at  low  densities
first by  respecting
 their properties at nuclear matter. Then,  the  extrapolations  of  the
 coupling constants in the RDH or RDHF approaches are restricted by the
 scalar and vector potentials at the low densities. Therefore, the
 extrapolations of coupling constants
 at the low densities in different cases are on an equal level.
 As mentioned above,   the  contributions  of isovector mesons  are neglected
 in Refs.\cite{bro92,fri93}. It is known that a realistic description of
 the nucleon-nucleon interaction in terms of meson exchange must include
 $\pi$ and $\rho$ . Therefore , it is necessary to develop a relativistic
 theory for finite nuclei in the RHF with the isovector mesons $\pi$ ,
 $\rho$ included. Fritz, M\"{u}ther \cite{fri94}
 discussed the $\pi$ contribution in the RDHF approach  to  the  bulk
properties  of  finite
 nuclei. They found that
 the inclusion of the $\pi$-exchange terms in the RDHF slightly
 improve the agreement between calculation and experiment. On  the  other
 hand,
 Boersma and Malfliet \cite{boe94} achieved a  density  dependent
parametrization
 of the Dirac-Brueckner G  matrix  in  nuclear  matter  which was called  an
 effective DBHF .
 They used the effective DBHF to  systematicly  analyze  a  series
 of spherical nuclei. The  results  are  in  good  agreement  with
 the experiments. It should be mentioned that the isovector vector meson $\rho$
was  not
 included in those calculations. In our  previous
 brief report \cite{ma94p}, the effects of the isovector $\rho$ meson in the
RDHF  on
 the bulk properties for finite nuclei are discussed  but   $\rho$
 tensor coupling was not included.  It was found that the Fock exchange terms
in
 the $\sigma$-$\omega$ model reduce the charge radii, but have less influence
on  the
 binding energies.  A  large  repulsion  of $\pi$ contribution  at  the
 interior of nuclear is observed. As a  result  the  energy levels  of
 single particles becomes shallow at the presence of $\pi$ meson.   Therefore
 the total binding energy is reduced and the charge radius is  expanded.
 This effect is partly canceled when the $\rho$ meson  is  included  in  the
 RDHF approach. But the contribution of tensor coupling of $\rho$ meson has
 not been discussed in these works .
 In this paper,   the  effect  of  tensor  coupling  of $\rho$ meson  is
 included in the RDHF approach. The systematic study of finite  nuclei
 in terms of the RDHF approach is investigated.
     \end{sloppypar}

    The arrangement of this paper is as follows:
   The general formalism in this work is presented
in Sec.2 . The numerical results and main conclusions are included in Sec.3
and Sec4.

     \begin{center}
     \section{The formalism}
     \end{center}
     \begin{sloppypar}

As in the one-boson-exchange(OBE) description of the NN
interaction\cite{mac89},our
starting point is an effective Lagrangian density which couples a nucleon
($\psi$) to two isoscalar mesons($\sigma$ and $\omega$) and two isovector
ones( $\pi$ and $\rho$) with the following quantum number ($J^{\pi}$,T):
$$\sigma(0^+,0),\omega(1^-,0),\pi(0^-,1),\rho(1^-,1)$$
and the electromagnetic field $(A^{\mu})$ is also included.\\

     \end{sloppypar}
     \begin{sloppypar}

   The effective Lagrangian density can be written as the sum of free and
interaction parts:
\begin{equation}
{\cal L}={\cal L}_{0}+{\cal L}_{I}
\end{equation}
The free Lagrangian density is given by
$${\cal L}_{0}=\bar{\psi}(i\gamma_{\mu}\partial^{\mu}-M)\psi+\frac{1}{2}(
\partial_{\mu}\sigma\partial^{\mu}\sigma-m_{\sigma}^{2})+\frac{1}{2}m_{\omega}^{2}\omega_{\mu}\omega^{\mu}$$
$$ - \frac{1}{4}F_{\mu \nu}F^{\mu \nu} + \frac{1}{2} m_{\rho}^{2}{\bf
\rho}_{\mu}\cdot {\bf \rho}^{\mu}-\frac{1}{4}
{\bf G}_{\mu \nu}\cdot {\bf G}^{\mu \nu}$$
\begin{equation}
+\frac{1}{2}(\partial_{\mu} {\bf \pi} \cdot \partial^{\mu}
{\bf \pi} - m_{\pi}^{2}{\bf \pi}^{2})- \frac{1}{4}H_{\mu \nu}H^{\mu \nu}  ,
\end{equation}
with
$$F_{\mu \nu}=\partial_{\nu}\omega_{\mu}-\partial_{\mu}\omega_{\nu}  ,$$
$${\bf G}_{\mu \nu}=\partial_{\nu}{\bf \rho}_{\mu}-\partial_{\mu}{\bf
\rho}_{\nu}  ,$$
$$H_{\mu \nu}=\partial_{\nu}A_{\mu}-\partial_{\mu}A_{\nu}   ,$$
where the meson fields are denoted by $\sigma$ , $\omega_{\mu}$ , ${\bf
\rho}_{\mu}$ and
${\bf \pi}$ , and $m_{\sigma}$ , $m_{\omega}$ , $m_{\rho}$ and $m_{\pi}$ are
their masses , respectively.
The nucleon field is denoted by $\psi$ which has a rest mass $M$ . $A_{\mu}$ is
the
electromagnetic field. The interaction Lagrangian density is given by
$${\cal L}_{I}=g_{\sigma} \bar{\psi} \sigma \psi -
g_{\omega}\bar{\psi}\gamma_{\mu}
\omega^{\mu}\psi - g_{\rho}\bar{\psi}\gamma_{\mu}{\bf \rho}^{\mu}\cdot {\bf
\tau}\psi $$
$$+\frac{f_{\rho}}{2M}\psi\sigma_{\mu \nu}\partial^{\mu}{\bf \rho}^{\nu}\cdot
{\bf \tau}\psi
- e\bar{\psi}\gamma_{\mu}\frac{1}{2}(1+\tau_{3})A^{\mu}\psi $$
\begin{equation}
-\frac{f_{\pi}}{m_{\pi}}\psi\gamma_{5}\gamma_{\mu}\partial^{\mu}{\bf \pi}\cdot
{\bf \tau}\psi  ,
\end{equation}
here ${\bf \tau}$ and $\tau_{3}$ are the usual isospin Pauli matrices. The
effective
strengths of couplings between the mesons and nucleons are denoted by the
coupling
constants $g_{i}$ or $f_{i}$ ($i=\sigma,\omega,\rho,\pi$) , respectively. Note
that
the pseudovector(PV) coupling for $\pi$NN interaction is used , because
the baryon self-energies are extremely large
(about 40 times larger than their PV counterpart) at normal nuclear density
if a pseudoscalar coupling is used , which has a drastic effect on the
single-particle spectrum\cite{hor83}. The present of tensor couplings makes the
model
Lagrangian density no longer renormalizable and all physical observable
should be calculated at the tree level \\

     \end{sloppypar}

\subsection{Equations of motion}

     \begin{sloppypar}

  The equation of motion for the meson fields are easily obtained from the
Euler-
Lagrange equation
\begin{equation}
\frac{\partial{\cal L}}{\partial\phi}-\partial_{\mu}\frac{\partial{\cal L}}
{\partial(\partial^{\mu}\phi)}=0  ,
\end{equation}
with a meson field $\phi$ . For instance , the $\sigma$ and $\omega$ fields are
the solutions of
\begin{equation}
(\Box+m_{\sigma}^{2})\sigma=g_{\sigma}\bar{\psi}\psi  ,
\end{equation}
\begin{equation}
(\Box+m_{\omega}^{2})\omega_{\nu}=g_{\omega}\bar{\psi}\gamma_{\nu}\psi  ,
\end{equation}
with the baryon current conservation $\partial^{\mu}(\bar{\psi}
\gamma_{\mu}\psi)=0$. Solving the equation for the meson fields , one then
obtains
\begin{equation}
\sigma(x)=g_{\sigma}\int d^{4}y D_{\sigma}(x-y) \bar{\psi}(y)\psi(y)  ,
\end{equation}
\begin{equation}
\omega^{\mu}(x)=g_{\omega}\int d^{4}y D_{\omega}^{\mu \nu}(x-y)
\bar{\psi}(y)\gamma_{\nu}\psi(y) ,
\end{equation}
where $D_{\sigma}(x-y)$ and $D_{\omega}^{\mu \nu}(x-y)$ is the $\sigma$ and
$\omega$
meson propagator. Similar expressions are deduced for isovector mesons.

     \end{sloppypar}

     \begin{sloppypar}

   Following standard techniques\cite{fet71}, at the Hartree-Fock level , the
average of
Hamiltonian at the ground state can be written as
$$<\Phi_{0}|H|\Phi_{0}>=\sum_{\alpha}\int U_{\alpha}^{\dagger}(x)[-i\alpha\cdot
\nabla+
\gamma_{0} m]U_{\alpha}(x) dx +\int
U_{\alpha}^{\dagger}(x)\gamma_{0}\Sigma_{H}(x)
U_{\alpha}(x) dx $$
\begin{equation}
 - \int U_{\alpha}^{\dagger}(x)\gamma_{0}\int\Sigma_{F}(x,y)U_{\alpha}(y)
dy dx   ,
\end{equation}
where $U_{\alpha}(x)$ is nucleon wave function and satisfies the orthogonality
relation
$$\int dy U_{\alpha}^{\dagger}(y) U_{\beta}(y)=\delta_{\alpha\beta}$$
Only positive-energy states have been taken into account in the preceding
derivation.
\\

     \end{sloppypar}

\subsection{Nuclear matter}

     \begin{sloppypar}

   Because of the translational and rotational invariance in the rest frame of
infinite nuclear matter and the assumed invariance under parity and time
reversal,
the nucleon self-energy produced by the meson exchanges in nuclear matter
can , in general, be written as
\begin{equation}
\Sigma(k_{\nu})=\Sigma_{s}(k_{\nu})-\gamma_{0}\Sigma_{0}(k_{\nu})+
{\bf \gamma} \cdot {\bf k}\Sigma_{v}(k_{\nu})
\end{equation}
where $\Sigma_{s}$ , $\Sigma_{0}$ , $\Sigma_{v}$ denote the scalar, time and
space components of vector potentials, respectively. In general, they are
functions of the
four-momentum $k_{\nu}$ of a nucleon and Fermi momentum $k_{F}$. Based on the
Feynman diagram rules one
could derive the nucleon self-energy in nuclear matter. In the RHF
approach, the isoscalar mesons in our Lagrangian density give
rise to the following contributions to the self-energy\cite{hor83,ma88}:
\begin{eqnarray}
\Sigma_{s}(k_{\nu})&=&-(\frac{g_{\sigma}}{m_{\sigma}})^{2}\rho_{s}+\frac{1}{16\pi^2k}
\int_{0}^{k_{F}}dq q \hat{M}(q_{0})[{g_{\sigma}^2\Theta_{\sigma}(k,q)-
4 g_{\omega}^2\Theta_{\omega}(k,q)}], \\
\Sigma_{0}(k_{\nu})&=&-(\frac{g_{\omega}}{m_{\omega}})^{2}\rho_{B}-\frac{1}{16\pi^2k}
\int_{0}^{k_{F}}dq q [{g_{\sigma}^2\Theta_{\sigma}(k,q)+
2 g_{\omega}^2\Theta_{\omega}(k,q)}], \\
\Sigma_{v}(k_{\nu})&=&-\frac{1}{(8 \pi^2 k^2)}\int_{0}^{k_{F}}dq q
\hat{Q}(q_{\nu})
[{g_{\sigma}^2\Phi_{\sigma}(k,q)+2 g_{\omega}^2\Phi_{\omega}(k,q)}]   ,
\end{eqnarray}
where
$$\hat{M}(q_{\nu})=\frac{M^{*}(q_{\nu})}{q_{0}^{*}(q_{\nu})},\hspace{1cm}
\hat{Q}(q_{\nu})=
\frac{q^{*}(q_{\nu})}{q_{0}^{*}(q_{\nu})}$$
$$\Theta_{i}(k,q)=ln\left|\frac{(k+q)^2+m_{i}^2}{(k-q)^2+m_{i}^2 }\right| $$
$$\Phi_{i}(k,q)=\frac{k^2+q^2+m_{i}^2}{4kq}\Theta_{i}(k,q) -1,\hspace{1cm}
i=\sigma,\omega,\rho,\pi$$
$${\bf k}^{*}={\bf k}(1+\Sigma_{v}(k_{\nu})),\hspace{1cm} k^{*}=|{\bf k}^{*}|$$
$$M^{*}=M+\Sigma_{s}(k_{\nu}),$$
$$k_{0}^{*}=k_{0}+\Sigma_{0}(k_{\nu})=({k^{*}}^2+{M^*}^2)^{1/2}   .$$
The scalar and vector densities are
$$\rho_{s}=\frac{2}{\pi^2}\int_{0}^{k_{F}} q^2 \hat{M}(q) dq,$$
\begin{equation}
\rho_{B}=\frac{2}{3 \pi^2} k_{F}^3
\end{equation}
The contributions of the isovector mesons $\pi$ and $\rho$ to the self-energy
is given in the Appendix A.
The first terms of the $\Sigma_{s}$ and $\Sigma_{0}$ are the Hartree terms,
which
are energy-independent. The rest terms are the Fock terms, which are almost 1/k
energy dependent. The coupled nonlinear integral equations (11-13) have to be
solved self-consistently . \\

With these nucleon self-energies , following the procedure discussed in Sec3.1,
we can obtain the coupling constants $g_{\sigma}$ and $g_{\omega}$ at each
baryon density.

     \end{sloppypar}

\subsection{ Finite nuclei}

     \begin{sloppypar}

  At the case of spherical, closed-subshell nuclei, a single-particle baryon
state with energy $E_{\alpha}$ is specified by the set of quantum numbers
$$\alpha=(q_{\alpha},n_{\alpha},l_{\alpha},j_{\alpha},m_{\alpha})\equiv(a,m_{\alpha})$$
where $q_{\alpha}=-1$ for a neutron state and $q_{\alpha}=+1$ for a proton
state.
The nucleon wave function can be written as
\begin{equation}
U_{\alpha}(x)=\frac{1}{r}\left(\begin{array}{c}
iG_a(r)\\
F_a(r){\bf \sigma} \cdot \hat{{\bf{r}}}
\end{array} \right)
{\cal Y}_{\alpha}\chi_{1/2}(q_{\alpha})  ,
\end{equation}
where $\chi_{1/2}(q_{\alpha})$ is an isospinor, and the angular and spin parts
of the nucleon spinor can be written as
$${\cal
Y}_{\alpha}(\hat{r})=\sum_{\mu_{\alpha},s_{\alpha}}<l_{\alpha}\frac{1}{2}
\mu_{\alpha} s_{\alpha}|j_{\alpha} m_{\alpha}>
Y_{l_{\alpha}}^{\mu_{\alpha}}(\hat{\bf{r}})
\chi_{1/2}(s_{\alpha})  . $$
The spinors $U_{\alpha}(r)$ are normalized according to
$$\int d^3 {\bf{r}}U_{\alpha}^{\dagger}(x)
U_{\alpha}(x)=\int_{0}^{\infty}[G_a^{2}(r)
+F_a^{2}(r)] dr=1  .$$
\\

     \end{sloppypar}
     \begin{sloppypar}

   The HF solution is obtained by requiring that the total binding energy
$$ E=<\phi_{0}|H|\phi_{0}>-AM,$$
is stationary with respect to variations of the spinors $U_{\alpha}$(i.e. of
$G_{\alpha}$ and
$F_{\alpha}$) such that the normalization relation is preserved
\begin{equation}
\delta[E-\sum_{\alpha(occ)} E_{\alpha}\int U_{\alpha}^{\dagger}(r)
U_{\alpha}(r)
d^3 r]=0
\end{equation}
using the expression of nucleon spinor, after a lengthy derivation, the HF
equations
for the self-consistent wave functions $(G_{\alpha},F_{\alpha})$ and energies
$E_{\alpha}$ will be obtained. The radial Dirac equation take the following
form:
$$\frac{d}{dr}\left(\begin{array}{c}
G_a(r)\\
F_a(r)
\end{array} \right)  =
\left(\begin{array}{cc}
-\frac{\kappa_{\alpha}}{r}-\Sigma_{T,a}^{D}(r)  &
M+E_{\alpha}+\Sigma_{S,a}^{D}(r)
-\Sigma_{0,a}^{D}\\
M-E_{\alpha}+\Sigma_{S,a}^{D}(r)+\Sigma_{0,a}^{D} &  \frac{\kappa_{\alpha}}{r}
+\Sigma_{T,a}^{D}(r)
\end{array} \right)
\left(\begin{array}{c}
G_a(r)\\
F_a(r)
\end{array}  \right) $$
\begin{equation}
+ \left(\begin{array}{c}
-X_{\alpha}(r)\\
Y_{\alpha}(r)
\end{array}  \right)   \label{eq-hf}
\end{equation}
here, $\Sigma_{S,a}^{D}$, $\Sigma_{0,a}^{D}$ and $\Sigma_{T,a}^{D}$ are direct
contributions to the self-energy and can be written as
\begin{equation}
\Sigma_{T,a}^{D}(r)=[\Sigma_{\rho}^{T}(r)+\Sigma_{\rho}^{VT(1)}(r)] q_{\alpha}
\end{equation}
\begin{equation}
\Sigma_{S,a}^{D}(r)=\Sigma_{\sigma}
\end{equation}
\begin{equation}
\Sigma_{0,a}^{D}(r)=\Sigma_{\omega}(r)+[\Sigma_{\rho}^{S}(r)+\Sigma_{\rho}^{VT(2)}(r)]
q_{\alpha} + \frac{1}{2}(1+q_{\alpha}) \Sigma_{c}(r)
\end{equation}

whereas $X_{\alpha}$ and $Y_{\alpha}$ come from exchange (Fock) contribution.
The quantity $\kappa_{\alpha}$ is $(2 j_{\alpha}+1)(l_{\alpha}-j_{\alpha})$.
\\
     \end{sloppypar}
     \begin{sloppypar}

    In this work we consider nuclei with a closed proton and neutron
shell only, therefore, the isovector pseudoscalar meson yield no contributions
in the Hartree approximation. The Hartree contributions come from
$\sigma$, $\omega$, $\rho$ mesons and Coulomb
force are given as
\begin{equation}
 \Sigma_{\sigma}(r)=-g_{\sigma}(\rho_B(r)) m_{\sigma} \int_0^{\infty}
g_{\sigma}
(\rho_B (r')) \rho_S (r') \tilde{I}_{0}(m_{\sigma}r_{<})
\tilde{K}_{0}(m_{\sigma}r_{>}) r'^2 dr' \\
\end{equation}
\begin{equation}
 \Sigma_{\omega}(r)=g_{\omega}(\rho_B(r)) m_{\omega} \int_0^{\infty} g_{\omega}
(\rho_B (r')) \rho_B (r') \tilde{I}_{0}(m_{\omega}r_{<})
\tilde{K}_{0}(m_{\omega}r_{>}) r'^2 dr' \\
\end{equation}
\begin{equation}
 \Sigma_{\rho}^{S}(r)=g_{\rho}^2 m_{\rho} \int_0^{\infty}
 [\rho_{B,p} (r') - \rho_{B,n} (r')] \tilde{I}_{0}(m_{\rho}r_{<})
\tilde{K}_{0}(m_{\rho}r_{>}) r'^2 dr'\\
\end{equation}
\begin{equation}
 \Sigma_{\rho}^{VT(1)}=-\frac{f_{\rho}}{2 m g_{\rho}} \frac{d}{dr}
\Sigma_{\rho}^{S}(r) \\
\end{equation}
\begin{equation}
 \Sigma_{\rho}^{VT(2)}(r)=-\frac{g_{\rho} f_{\rho}}{2 M} m_{\rho}
\int_0^{\infty}
[\rho_{T,p} (r') - \rho_{T,n} (r')] [\frac{d}{dr}\tilde{I}_{0}(m_{\rho}r_{<})
\tilde{K}_{0}(m_{\rho}r_{>})] r'^2 dr'\\
\end{equation}
$$ \Sigma_{\rho}^{T}(r)=-(\frac{f_{\rho}}{2M})^2 \{ m_{\rho}^3 \int_0^{\infty}
[\rho_{T,p} (r') - \rho_{T,n} (r')] \tilde{I}_{1}(m_{\rho}r_{<})
\tilde{K}_{1}(m_{\rho}r_{>}) r'^2 dr'$$
\begin{equation}
-[\rho_{T,p} (r) - \rho_{T,n} (r)] \}
\end{equation}

with the definitions
$$\rho_{S,n or p}=\frac{1}{4 \pi r^2} \sum_{b(n or p)} \hat{j}_{b}^2
[G_{b}^2 (r)-F_{b}^2 (r)] $$

$$\rho_{B,n or p}=\frac{1}{4 \pi r^2} \sum_{b(n or p)} \hat{j}_{b}^2
[G_{b}^2 (r)+F_{b}^2 (r)] $$

$$\rho_{T,n or p}=\frac{1}{4 \pi r^2} \sum_{b(n or p)} \hat{j}_{b}^2
[2 G_{b} (r) F_{b} (r)] $$
and $r_{<}$ ($r_{>}$) is the smaller (larger) of the $r'$ and $r$.

The functions $\tilde{I}_{L}(x)$ and $\tilde{K}_{L}(x)$ arise from the multiple
expansion of the meson propagator in the coordinate space and are defined using
the modified spherical Bessel functions of the first and third kind I and K:
$$\tilde{I}_{L}(x)=\frac{I_{L+\frac{1}{2}}(x)}{\sqrt{x}},\hspace{1cm}
  \tilde{K}_{L}(x)=\frac{K_{L+\frac{1}{2}}(x)}{\sqrt{x}}$$
The explicit expressions of exchange contributions $X_{\alpha}$ and
$Y_{\alpha}$
are given in the Appendix B.\\

     \end{sloppypar}

     \begin{center}
     \section{Results and Discussions}
     \end{center}
\subsection{Parameterization of DBHF}

The self-energy obtained in the RHF is of very weak energy-dependence and
too strong density-dependence due to the fact that the short range
correlation has not been considered . Attempts to incorporate
the effects of the short range correlation described in the DBHF approach
have been made by introducing density- and momentum-dependent effective
coupling constants of mesons in the RHF approach .
In order to make comparison with the DBHF results, the scalar and
vector potentials can be obtained by
\begin{equation}
U_{s}(k,k_{F}) = \frac{\Sigma_{s} - M \Sigma_{v}}{1+ \Sigma_{v}},\; \;
 \; \;U_{o}(k,k_{F})=\frac{-\Sigma_{o}+E_{k}\Sigma_{v}}{1+\Sigma_{v}},
\end{equation}
where $E_{k} = \sqrt{{k^{\ast}}^{2}+{M^{\ast}}^{2}} - \Sigma_{o}$. The
momentum dependence of the potentials  usually are relatively weak and
neglected
in the description of ground state properties of finite nuclei .
Therefore , the momentum average within the Fermi sea is performed,
\begin{equation}
U_{s(o)}(k_{F}) = \frac{\int_{0}^{k_{F}} k^{2}U_{s(o)}(k,k_{F})~dk}
{\int_{0}^{k_{F}} k^{2}~dk}.
\end{equation}

     \begin{sloppypar}

At very low density of nuclear matter the DBHF results are not reliable
and remain unknown. Therefore, the extrapolation of the coupling constants
outside the density points in the DBHF has to be done when they are applied
to the calculation of finite nuclei. In order to remove the
sensitivity, the extrapolation of scalar and vector potentials U$_{s}$
and U$_{o}$ of the DBHF results at low densities are done by setting
U$_{s}$ = 0, U$_{o}$ = 0 at $\rho$ = 0. It is known that the scalar
and vector potentials in RMF or RHF are almost linear dependent on the
density. Due to the two-body correlation the scalar and vector
potentials  approach to zero smoothly as the density goes to zero.
A polynomial fit of the scalar and vector potentials with respect to
the density are performed. The extrapolation  and interpolation of U$_{s}$ and
U$_{o}$
are shown in Fig.1, where the circles are the DBHF results in nuclear
matter using the Bonn A potential\cite{bro84}. The density dependence
of the coupling constants are then adjusted in the cases of RMF or RHF
with or without isovector mesons to reproduce the nucleon
self-energies at each densities resulting from the DBHF.
The nucleon and $\sigma$ and $\omega$ meson
masses are chosen to be the same as the DBHF calculation, where
M = 938.9 MeV, m$_{\sigma}$ = 550 MeV, m$_{\omega}$ = 782.6 MeV. The
pseudo-vector coupling for $\pi$NN and vector and tensor coupling for $\rho$NN
are adopted. The masses and coupling constants of isovector mesons
are fixed to be m$_{\pi}$ = 138 MeV, m$_{\rho}$ = 770 MeV,
$\frac{f_{\pi}^{2}}{4\pi}$ = 0.08,
$\frac{g_{\rho}^{2}}{4\pi}$ = 0.55 , $\frac{f_\rho}{g_\rho}$ = 3.7
\cite{bou87}.
The density dependence of coupling constants in different cases are
shown in Fig.2. The presence of the pion introduces a large repulsive
force, so the scalar coupling constant becomes larger and the vector
coupling constant gets smaller to balance the repulsive  force,
especially at normal and high densities. However, the pion
contribution is partly canceled by the presence of vector part of
$\rho$ meson at the symmetric nuclear matter . The tensor coupling of $\rho$
meson has large effect at high density . As a result , the coupling constant
of $\sigma$ meson becomes larger than that of $\omega$ meson at high densities
{}.
The results obtained in this paper
are somewhat different from those in ref.\cite{fri94}. The reason
for this discrepancy is that the zero-range components of the
pion-exchange were removed there. The cases with and without the
contact interactions have been discussed in more detail in ref.\cite{ber93}.
It is found in our calculations that the effects of the contact interactions
mainly cause a renormalization in $g_\sigma$ and $g_\omega$ coupling
constants. The removal of the zero-range components of the pion and
rho exchange would increase the binding energy and reduce the charge
r.m.s radius. No qualitative improvement has been found. Therefore, only
the cases with the zero-range components of the $\pi$ and $\rho$
exchange are presented in this paper. \\
     \end{sloppypar}
\subsection{The ground state properties of finite nuclei}

The ground state properties of four stable doubly closed-shell nuclei
$^{16}$O, $^{40}$Ca, $^{48}$Ca and $^{90}$Zr are calculated with these
density dependent coupling constants in the RDH and RDHF approaches . The
set of coupled differential equations(17) is solved in the coordinate space
following the method of Ref.\cite{bou87}. The self-consistence is achieved
by an iterative procedure. It is different from a matrix diagonalization method
adopted in Refs.\cite{bro92,fri93}, where a termination of a complete
set of bases has been performed both for baryon and meson results.
Our computer code has been carefully checked with the results of
ref.\cite{bou87,fri-em} .

In order to investigate the effect of the density-dependence, the results for
$^{16}$O and $^{40}$Ca obtained in the RMF and RHF with
of $\sigma$ + $\omega$ (RHF1) , $\sigma$ + $\omega$ + $\pi$ (RHF2) and $\sigma$
+ $\omega$+ $\pi$+ $\rho$(RHF3) are list in Table 1. The coupling constants
are determinated to reproduce the DBHF results in nuclear matter (OBE potential
A)
at saturation density K$_{F}$ = 1.40 fm$^{-1}$.
It should be mentioned that the results are different from those of the usual
RMF and RHF calculations , where the coupling constants and scalar meson mass
are adjusted to reproduce the empirical saturation properties of nuclear
matter as well as the rms charge radius of $^{40}$Ca . Because of relative
large saturation density obtained in the DBHF and the scalar meson mass
$m_{\sigma}$ = 550 Mev adopted in this calculation , the binding energies
and rms charge radii calculated here are both much smaller than the
experimental data . However , the main purpose of Table 1 is to show the
difference in various cases mentioned above as well as the effect of the
density-dependence in comparison with Table 2.
The calculations with density-dependent effective interaction
are performed, where the coupling constants at each baryon density come from
the
parametrization of the DBHF result in nuclear matter as discussed in section
3.1.
Various cases, RDH, RDHF with $\sigma$ + $\omega$(RDHF1), $\sigma$ + $\omega$
+$\pi$ (RDHF2) and $\sigma$ +$\omega$ + $\pi$ + $\rho$ (RDHF3) are investigated
and the results
for the nuclei $^{16}$O,$^{40}$Ca , $^{48}$Ca and $^{90}$Zr are displayed in
Table 2 . In order to investigate the sensitivity of the $\rho$ meson
coupling constant , two values of $\rho$ coupling
constants are adopted in the calculations :
$\frac{g_\rho^2}{4\pi}$ = 0.55(RDHF3A) and 0.99(RDHF3B) without tensor
coupling.
The results with $\rho$NN tensor coupling ($\frac{g_\rho^2}{4\pi}$ = 0.55,
$\frac{f_\rho}{g_\rho}$ = 3.7) are given as RDHF3C.
\begin{table}
\begin{tabular}{cccccc}
\hline
& RMF & RHF1 & RHF2 & RHF3 & \\
\hline
\multicolumn{2}{c}{$^{16}$O} & & & \\
$E_{B}/A$(Mev)    &  -5.62 &  -6.02 &  -4.86  &  -5.67   \\
$r_{ch}$(fm)      &   2.48 &   2.39 &   2.53  &   2.57   \\
\hline
$1s_{1/2}$(Mev)   &  44.34 &  45.08 &  40.78  &  43.21   \\
$1p_{3/2}$(Mev)   &  18.96 &  21.15 &  17.50  &  18.51   \\
$1p_{1/2}$(Mev)   &   9.62 &   7.85 &   9.49  &  10.98   \\
\hline
\multicolumn{2}{c}{$^{40}$Ca} & & &\\
$E_{B}/A$(Mev)    & -6.36  &  -6.69 &  -5.84  &  -6.38   \\
$r_{ch}$(fm)      &  3.14  &   3.04 &   3.16  &   3.22   \\
\hline
$1d_{5/2}$(Mev)   & 16.54  &  18.61 &  15.72  &  16.35    \\
$2s_{1/2}$(Mev)   &  7.07  &   5.60 &   7.98  &   8.81    \\
$1d_{3/2}$(Mev)   &  6.92  &   5.70 &   6.50  &   7.79    \\
\hline
\end{tabular}
\caption{Ground state properties of $^{16}$O and $^{40}$Ca calculated by RHA
and RHF.
The binding energy per nucleon $E_{B}$/A, the charge rms radius r$_{c}$
and single-particle energies of proton states.
}
\end{table}

\begin{table}
\begin{tabular}{cccccccc}
\hline
& RDH & RDHF1 & RDHF2 & RDHF3A & RDHF3B & RDHF3C & Exp.\\
\hline
\multicolumn{2}{c}{$^{16}$O} & & & & & &\\
$E_{B}/A$(Mev)    &  -7.44  & -7.48  & -6.96   & -7.29  & -7.57 & -7.41 &
-7.98\\
$r_{ch}$(fm)      &   2.59  &  2.50  &  2.64   &  2.61  &  2.59 &  2.68 &
2.73\\
\hline
$1s_{1/2}$(Mev)   &  43.97  & 43.87  & 41.11   & 42.76  & 44.13 &  42.98 & 40
$\pm$ 8\\
$1p_{3/2}$(Mev)   &  21.77  & 23.60  & 21.11   & 21.85  & 22.47 &  21.31 & 18.4
\\
$1p_{1/2}$(Mev)   &  16.16  & 16.08  & 15.60   & 16.07  & 16.47 &  15.72 & 12.1
\\
\hline
\multicolumn{2}{c}{$^{40}$Ca} & & & & & &\\
$E_{B}/A$(Mev)    &   -7.88 & -7.91  & -7.49   &  -7.74 & -7.94  & -7.81 &
-8.55 \\
$r_{ch}$(fm)      &    3.26 &  3.17  &  3.29   &   3.27 &  3.26  &  3.35 &
3.48 \\
\hline
$1d_{5/2}$(Mev)   &   19.27 & 21.17  & 19.08   &  19.52 & 19.89  &  18.95 &
15.5 \\
$2s_{1/2}$(Mev)   &   13.69 & 14.08  & 14.20   &  14.19 & 14.14  &  13.78 &
10.9 \\
$1d_{3/2}$(Mev)   &   13.29 & 13.48  & 12.77   &  13.09 & 13.35  &  12.67 &
8.3 \\
\hline
\multicolumn{2}{c}{$^{48}$Ca} & & & & & &\\
$E_{B}/A$(Mev)    &   -8.02 & -7.96  & -7.45   &  -7.59 & -7.70  &  -7.60 &
-8.67 \\
$r_{ch}$(fm)      &    3.27 &  3.17  &  3.30   &   3.28 &  3.27  &   3.37 &
3.47 \\
\hline
$1d_{5/2}$(Mev)   &   24.35 & 29.17  & 24.34   &  26.37 & 27.96  &  26.72 &
20.0 \\
$2s_{1/2}$(Mev)   &   17.25 & 19.77  & 19.79   &  20.90 & 21.77  &  20.61 &
15.8 \\
$1d_{3/2}$(Mev)   &   18.68 & 21.79  & 23.05   &  24.92 & 26.38  &  23.53 &
15.3 \\
\hline
\multicolumn{2}{c}{$^{90}$Zr} & & & & & &\\
$E_{B}/A$(Mev)    &  -7.94  & -7.92  & -7.55   & -7.67  & -7.78  &  -7.68 &
-8.71 \\
$r_{ch}$(fm)      &   4.00  &  3.89  &  4.02   &  4.00  &  3.99  &   4.10 &
4.27 \\
\hline
$2p_{3/2}$(Mev)   &  11.01  & 13.09  & 12.86   & 13.61  & 14.19  &  13.52 &
11.0 \\
$1f_{5/2}$(Mev)   &  13.14  & 15.61  & 16.67   & 18.12  & 19.24  &  16.93 &
12.3 \\
$2p_{1/2}$(Mev)   &   9.42  & 11.16  & 11.77   & 12.49  & 13.05  &  11.91 &
9.5 \\
\hline
\end{tabular}
\caption{Ground state properties of $^{16}$O , $^{40}$Ca , $^{48}$Ca and
$^{90}$Zr calculated by RDH and RDHF.
The binding energy per nucleon $E_{B}$/A, the charge rms radius r$_{c}$
and single-particle energies of proton states. }
\end{table}

The importance of the density dependent approaches is clearly demonstrated
in Table 1 and Table 2. The calculations in either RMF or RHF with constant
coupling constants produce much smaller binding energies of nucleon
in comparison with the experiments. In contrast, the density-dependent
interactions increase both binding energy and charge radius, which
imply the removal from the so-called Coester band\cite{coe70}. The
results in the relativistic density-dependent approaches are
largely improved and closer to the experimental values. The slight
differences from ref.\cite{bro92,fri93} are due to the different
extrapolation procedures. The Fork exchange term in the $\sigma$-$\omega$
model reduces the charge radii, but has less influence on
the binding energies. A large repulsion of pion contribution at the
interior of nucleus is found. As a result, the energy levels of
single particles become shallow at the presence of pion. Therefore,
the total binding energy is reduced and the charge radius is
expanded. This effect is partly canceled by the $\rho$ meson
exchange contribution.  The charge radii of nuclei calculated in the RDHF
with isovector mesons are much close to those
obtained in the RDH, which can also be observed in the charge density
distributions.
In comparison of the RDHF3A and RDHF3B , the results are not sensitive to the
strength of the $\rho$ meson coupling. The binding energies for the
strong coupling constant of $\rho$ meson $\frac{g_\rho^2}{4\pi}$ = 0.99
is about 2\% bigger than those for $\frac{g_\rho^2}{4\pi}$ = 0.55
and the rms radius is reduced less than 1\%. With $\rho$ tensor coupling
,it can be found that the results of both binding energy and rms charge
radius are improved . The binding energies of nuclei is increased slightly, but
the
charge radii of nuclei are improved largely in comparison with experiments.

Figure 3 shows the density distribution of nuclei. The dash-dotted
curves are the results of the RHF3, corresponding to the fourth column  in
Table 1. The dashed, dotted and solid ones correspond to those of the
RDH and RDHF without and with isovector mesons (RDHF1, RDHF3A, RDHF3C) ,
respectively .
It is found that the
charge densities are reduced at the nuclear interior and have long
tail  due to the relative strong coupling constants at the nuclear surface
in the density dependent calculations. The Fock contribution of
 $\sigma$ and $\omega$ in the RDHF produce a squeezing effect and
give a large central density. The repulsive contribution of $\pi$
reduces the interior density and the results of the
RDHF3A are very close to those of the RDH . Though the
density at center is still higher than the experimental results, it
is a parameter-free calculation in the sense that no parameters are
adjusted for the calculations of the many-body problem. However, the
results with density dependent coupling constants are in a
reasonable good agreement with the experiments.

As well known , nonrelativistic BHF calculations with various tow-body
nucleon-nucleon potentials, such as Reid
soft-core ,Hamada-Johnson potentials , reveal a "Coester" band in dependence
of $1/r_{c}$ on $E_{B}/A$. In the RMF or RHF calculations, the dependence of
$1/r_{c}$
on $E_B/A$ at a fix $K_{F}$, with variation of the scalar meson mass and
therefore the variation of the coupling constants , formed a new "Coester"
band. A better estimate of the merits of the present work can be expected
upon the comparison with the BHF "Coester" band and RHF "Coester" band. Those
"Coester" bands are plot in Fig.4 for $^{16}$O and $^{40}$Ca , the dash-dotted
line indicates the BHF "Coester"
band, which is taken from a so called generalized BHF calculations by K{\"
u}mmel
et al.\cite{kum78}. The dotted and dashed lines represent the RHF results ,
which are
obtained in the RHF3 by varying the scalar meson mass as well as the
different coupling constants to reproduce the
nuclear matter saturation properties at $K_{F}$ = 1.40fm$^{-1}$ (1) and
$K_{F}$ = 1.30fm$^{-1}$ (2) resulting from
the DBHF approach , respectively . The results in the case of the RDHF3C are
displayed by a solid line in the figure. The density dependent approach forms a
new
line away from all of the "Coester" band of conventional BHF and
seems to be much closer to the experimental values than the
BHF and RHF calculations.

The spin-orbit splittings of nuclei are given in Table 3. It can be seen
that the spin-orbit splitting in the RMF and RHF approaches is  larger
than the experimental data, which indicates the larger spin-orbit
force. It is known that the spin-orbit force is related to the derivative of
the
potentials with respect to the space and is a surface effect. The density
dependent
approaches reduce the sharp surface and, therefore, reduce the spin-orbit
splitting. The Fock terms of the $\sigma$ and $\omega$ exchange increase
the spin-orbit splitting, while the $\pi$ and $\rho$ exchanges give the
opposite contribution. A large reduction of the spin-orbit splitting
due to pion-exchange is found, especially for heavy nuclei with large
neutron excess. A flip of the spin-orbit splitting in $^{208}$Pb is
observed in the calculation of RDHF with all mesons included, which
is certainly not physical. It might indicate that the free coupling
constants of the isovector mesons adopted in the RDHF are too strong
for the nuclear structure calculation. A similar observation was obtained
in the calculation of the relativistic optical potential in the RHF
approach\cite{ma88}. The density dependent coupling constants for
isovector mesons  may also be required.

\begin{table}
\begin{tabular}{ccccccccc}
\hline
Nuclei     & RMF   & RHF1   &   RHF3 & RDH  & RDHF1 & RDHF3A & RDHF3C & Exp.\\
\hline
$^{16}$O   & 9.34  & 13.30  & 7.53 & 5.61 & 7.52 & 5.78 & 5.60 & 6.3 \\
$^{40}$Ca  & 9.62  & 12.82  & 8.56 & 5.78 & 7.69 & 6.43 & 6.28 & 7.2 \\
$^{48}$Ca  & 9.32  & 12.52  & 5.32 & 5.67 & 7.37 & 1.45 & 3.19 & 4.3 \\
\hline
\end{tabular}
\caption{Spin-orbit splittings of protons for the $1p$ shell in $^{16}$O and
the $1d$ shell in $^{40}$Ca and $^{48}$Ca.
}
\end{table}

It is known that the single particle densities are not directly provided
by experiments . The only way to gain some insight in the single particle
distribution is to
study the difference between density distribution of nearby nuclei. In Fig.5
,we give the charge distribution difference between $^{40}$Ca and $^{48}$Ca,
multiplied by $r^{2}$ . The difference of the neutron densities between
$^{40}$Ca and $^{48}$Ca are plotted in Fig.6 . The shaded area presents the
experimental data, the dashed curve is the results of the RDH ,
the dotted one the results of RDHF1 , and the solid one
is obtained in the case of RDHF3C. It can be seen that the results obtained
in the RDHF3C are superior to those of the RDH and RDHF1 in comparison with
experiments.
It means that the isospin dependence can not be correctly described by the RDH
as well as RDHF1 without including isovector mesons.

Neutron skin thickness is an important quantity to study isotope shifts.
It is defined as the difference between neutron and proton rms radii:
$\Delta_{np}=r_{n}-r_{p}$. The neutron skin thickness of $^{16}$O , $^{40}$Ca ,
$^{48}$Ca and $^{90}$Zr are given in Table 4 , and the $\Delta_{np}$ versus the
asymmetry parameter
(N-Z)/A for $^{40}$Ca, $^{48}$Ca and $^{90}$Zr are shown in Fig.7 .
The results of the RDHF3C seem to be similar to those of the RMF , which
are different from what obtained in ref.\cite{boe94} . More information of
of isospin dependence  the ground state properties are required .
\begin{table}
\begin{tabular}{cccccc}
\hline
Nuclei     & RHF3  &  DBHF   & RDHF1 & RDHF3C & Exp.\\
\hline
$^{16}$O   & -0.03  & -0.03  & -0.03 & -0.02 & -0.02 \\
$^{40}$Ca  & -0.05  & -0.06  & -0.05 & -0.04 & -0.07 - 0.10 \\
$^{48}$Ca  &  0.23  &  0.13  &  0.15 &  0.22 & 0.16 - 0.23 \\
$^{90}$Zr  &  0.11  &  0.04  &  0.07 &  0.11 &  0.07 \\
\hline
\end{tabular}
\caption{Neutron skin thickness $\Delta_{np}=r_n - r_p $ for $^{16}$O ,
 $^{40}$Ca , $^{48}$Ca , $^{90}$Zr . The DBHF results are taken from
[20].}
\end{table}

\section{Conclusion}

   In summary , the RDH and RDHF approaches with the density-dependent
effective coupling constants of isoscalar mesons can incorporate the DBHF
results and contain the nucleon-nucleon correlation effects . Inclusion
of the NN correlation let to a substantical improvement in the microscopic
description of bulk properties of nuclei . The Fork exchange terms are not
negligible , though the exchange contributions are relatively weak than
those of the Hartree direct term in the relativistic approach and their
contribution to the binding energy may be compensated by the variation
of the coupling constants . The important contributions from the isovector
meson $\pi$ and as well as , to some extent , $\rho$ meson are not included
in the mean field approach . It is found that the isovector mesons $\pi$
and $\rho$ play an important role in the spin-orbit splitting as well as the
isospin dependent quantities . The tensor coupling of $\rho$ meson gives a
constructive contribution to the bind energy , especially the rms charge
radius , therefore improves the agreement with the experimental data . More
information of the isospin denpendence of nuclear properies is required to
provide constraints on the coupling constants of isovector mesons $\pi$ and
$\rho$ in the nuclear medium . \\

\Large
\begin{center}
{\bf{Acknowledgment}}
\end{center}
\normalsize

This work was supported by The National Natural Science
Foundation of China.

\newpage
\large
\begin{center}
{\bf{Appendix A.{\ }Self-Energy in Nuclear Matter}}
\end{center}
\normalsize

    Base on Hartree-Fork approach , the nucleon self-energy in nuclear matter
coming from the contributions of isovector pseudoscalar meson $\pi$ can be
written
as follow:
$$
\begin{array}{l}
\Sigma_s^{\pi}(k)=\frac{3}{8\pi^2 k} \int_0^{k_{F}} dq q
(\frac{f_{\pi}}{m_{\pi}})^2
\hat{M} ( 2 k q - \frac{1}{2} m_{\pi}^2 \Theta_{\pi} )
\end{array}
\eqno{(A.1)}
$$
$$
\begin{array}{l}
\Sigma_0^{\pi}(k)=\frac{3}{8\pi^2 k} \int_0^{k_{F}} dq q
(\frac{f_{\pi}}{m_{\pi}})^2
[ -\frac{1}{2} m_{\pi}^2 \Theta_{\pi} + 2 k q ]
\end{array}
\eqno{(A.2)}
$$
$$
\begin{array}{l}
\Sigma_v^{\pi}(k)=-\frac{3}{8\pi^2 k^2} \int_0^{k_{F}} dq q
(\frac{f_{\pi}}{m_{\pi}})^2
[  q \hat{Q} k \Theta_{\pi} -\hat{Q} ( k^2 +q^2 ) \Phi_{\pi} ]
\end{array}
\eqno{(A.3)}
$$
The contribution coming from isovector vector meson $\rho$ can be write as :
$$
\begin{array}{c}
\Sigma_s^{\rho}(k)=\frac{3}{8\pi^2 k} \int_0^{k_{F}} dq q \{ -2 g_{\rho}^2
\hat{M} \Theta_{\rho}
+3 (\frac{f_{\rho}}{2 M})^2 \hat{M} ( 2 k q -\frac{1}{2} m_{\rho}^2
\Theta_{\rho} ) \\
+3 g_{\rho} (\frac{f_{\rho}}{2 M}) [q \hat{Q} \Theta_{\rho} - 2 k \hat{Q}
\Phi_{\rho}] \}
\end{array}
\eqno{(A.4)}
$$
$$
\begin{array}{l}
\Sigma_0^{\rho}(k)=\frac{3}{8\pi^2 k} \int_0^{k_{F}} dq q \{ g_{\rho}^2
\Theta_{\rho}
+(\frac{f_{\rho}}{2 M})^2 [ 2 k q -\frac{1}{2} m_{\rho}^2 \Theta_\rho ] \}
\end{array}
\eqno{(A.5)}
$$
$$
\begin{array}{c}
\Sigma_v^{\rho}(k)= - \frac{3}{8\pi^2 k^2} \int_0^{k_{F}} dq q \{ 2 g_{\rho}^2
\hat{Q} \Phi_{\rho}
+ 2 (\frac{f_{\rho}}{2 M})^2 [ k q \hat{Q} \Theta_{\rho} - \hat{Q} ( k^2 + q^2
-\frac{1}{2} m_{\rho}^2 ) \Phi_{\rho} ] \\
 -  3 g_{\rho} (\frac{f_{\rho}}{2 M}) ( k \hat{M} \Theta_{\rho} - 2 q \hat{M}
\Phi_{\rho} ) \}
\end{array}
\eqno{(A.6)}
$$

\large
\begin{center}
{\bf{Appendix B.{\ }Fock Term Expressions}}
\end{center}
\normalsize

   The quantities $X$ and $Y$ of eq.[\ref{eq-hf}] can be written as the sum of
contributions coming from different mesons . In the following , we often need
the reduced matrix elements of the tensorial operators $Y_{L}^m (\hat{r})$ and
$$ T_{JL}^{M} \equiv \sum_{m k} <L1mk|JM> Y_L^m(\hat{r}) \sigma^k $$
They are given as follow:
$$
\begin{array}{l}
<a||Y_L||b> = \left\{
\begin{array}{ll}
(4 \pi)^{-\frac{1}{2}} \hat{j}_a \hat{j}_{b} \hat{L} (-1)^{j_b-L-\frac{1}{2}}
\left( \begin{array}{lll}
j_a & j_b & L \\
\frac{1}{2} & -\frac{1}{2} & 0
\end{array}
\right)   & {\rm if\ } l_a+l_b+L{\ } is {\ }  even   \\
0 & {\rm if\ } l_a+l_b+L{\ } is {\ } odd
\end{array}
\right. \\
<a||T_{JL}||b> = (\frac{6}{4 \pi})^{\frac{1}{2}} (-1)^{l_a} \hat{j}_a
\hat{j}_{b} \hat{l}_{a} \hat{l}_b \hat{J} \hat{L} \left(
\begin{array}{lll}
l_a & L & l_b \\
0   & 0 & 0
\end{array} \right)
\left \{ \begin{array}{lll}
j_a & j_b & J \\
l_a & l_b & L \\
\frac{1}{2} & \frac{1}{2} & 1
\end{array}
\right \} {\ },
\end{array} \eqno{(B.1)}
$$
where $\hat{J} = \sqrt{2J+1} $.

The contribution coming from scalar moson $\sigma$ is determined to be
\begin{eqnarray*}
\left( \begin{array}{l}
 -X^{\sigma}(r) \\
   Y^{\sigma}(r)
\end{array} \right)
& = & g_{\sigma}(\rho_B(r)) m_{\sigma} \hat{j}_a^{-2} \sum_b \delta_{q_a q_b}
\left( \begin{array}{l}
 F_b(r)  \\
   G_b(r)
\end{array} \right)
\sum_L |<a||Y_L||b>|^2 \\
&\times& \int_0^{\infty} g_{\sigma}(\rho_B(r')) [G_a G_b - F_a F_b]_{r'}
\tilde{I}_L(m_{\sigma} r_<) \tilde{K}_L(m_{\sigma} r_>) d r' ~~,~~~~~~~~~(B.2)
\end{eqnarray*}
the sum over b running over occupied states .

The expressions for the vector meson $\omega$ are splited into timelike and
spacelike parts , due to the respective $\gamma_0$ and ${\bf{\gamma}}$
coupling.
The time component is
\begin{eqnarray*}
\left( \begin{array}{l}
 -X^{\omega}_0(r) \\
   Y^{\omega}_0(r)
\end{array} \right)
& = & g_{\omega}(\rho_B(r)) m_{\omega} \hat{j}_a^{-2} \sum_b \delta_{q_a q_b}
\left( \begin{array}{l}
 F_b(r)  \\
  - G_b(r)
\end{array} \right)
\sum_L |<a||Y_L||b>|^2 \\
&\times& \int_0^{\infty} g_{\omega}(\rho_B(r')) [G_a G_b + F_a F_b]_{r'}
\tilde{I}_L(m_{\omega} r_<) \tilde{K}_L(m_{\omega} r_>) d r'~~,~~~~~~~~~(B.3)
\end{eqnarray*}
The space component is
\begin{eqnarray*}
\left( \begin{array}{l}
 -X^{\omega}(r) \\
   Y^{\omega}(r)
\end{array} \right)
& = & - \frac{g_\omega (\rho_B(r))}{4 \pi } m_\omega \sum_{b,L} \delta_{q_a
q_b}
(2 j_b+1)(2 L+1)
\left( \begin{array}{l}
G_b(r) \\
-F_b(r)
\end{array} \right) \\
&\times& \int_0^{\infty} g_{\omega}(\rho_B(r'))  \{
\left( \begin{array}{l}
G_a F_b \\
-F_a G_b
\end{array} \right)
\left( \begin{array}{lll}
j_a & j_b & L \\
\frac{1}{2} & -\frac{1}{2} & 0
\end{array} \right) ^2  ~~~~~~~~~~~~~~~~~~~~~~~~(B.4)  \\
& + & \left( \begin{array}{l}
F_a G_b \\
G_a F_b
\end{array} \right)
[ 2 \left( \begin{array}{lll}
l_a & L & l_b'  \\
0   & 0 &  0
\end{array} \right) ^2
- \left( \begin{array}{lll}
j_a  &  j_b  &  L  \\
\frac{1}{2} & -\frac{1}{2} & 0
\end{array} \right) ^2
] \}  \\
&\times& \tilde{I}_L(m_{\omega} r_<) \tilde{K}_L(m_{\omega} r_>) d r'
\end{eqnarray*}
where $ a' = ( q_a,n_a,l_a',j_a) $ with $ l_a'=2 j_a - l_a $ .

The contribution comes from isovector pseudoscalar meson $\pi$ with
pseudovector
coupling is as follows
\begin{eqnarray*}
\left( \begin{array}{l}
 -X^{\pi} \\
  Y^{\pi}
\end{array} \right)
& = & f_{\pi}^2 \hat{j}_a^{-2} \sum_b (2-\delta_{q_a q_b}) \{  \frac{j_a^2
j_b^2}{8 \pi}
\frac{(G_a G_b + F_a F_b)_r}{m_{\pi}^2 r^2}
\left( \begin{array}{l}
 -F_b(r) \\
  G_b(r)
\end{array} \right) \\
& - & m_{\pi} \sum_L \hat{L}^{-4} |<a||Y_L||b'>|^2 \sum_{L_1,L_2}
\left( \begin{array}{l}
F_b(r)[\kappa_{ab}-\alpha(L_1)]  \\
G_b(r)[\kappa_{ab}+\alpha(L_1)]
\end{array} \right) i^{L_2-l_1} ~~~(B.5) \\
&\times& \int_0^{\infty} [\{ \kappa_{ab} + \alpha(L_2) \} G_a G_b -
\{ \kappa_{ab} - \alpha(L_2) \} F_a F_b ]_{r'} R_{L_1,L_2}(m_\pi r , m_\pi r')
dr' \}  ~~ .
\end{eqnarray*}
where we introduced the notation
$$ \kappa_{ab} = \kappa_a + \kappa_b, $$
$$ \alpha(L_1) = \left \{
\begin{array}{ll}
 -L & {\rm if\ } L_1=L-1, \\
 L+1 & {\rm if\ } L_1=L+1,
\end{array} \right.   $$
$$ R_{L_1 L_2} (m r , m r') = \tilde{I}_{L_1}(mr) \tilde{K}_{L_2} (m r')
\theta(r'-r)
+ \tilde{K}_{L_1}(mr) \tilde{I}_{L_2}(mr') \theta(r-r').  $$
The $L_1$ and $L_2$ can only take two values $L+1$ or $L-1$ .

The vector part of the $\rho$NN coupling in our Lagrangian gives $X(r)$ and
$Y(r)$ which are formally idential to those of the $\omega$ meson , expect for
the isospin factor $\delta_{q_a q_b}$ replaced by 2 - $\delta_{q_a q_b}$
and $m_{\omega}$ , $g_{\omega}(\rho_B(r))$ replaced by $m_{\rho}$ , $g_{\rho}$
{}.
The tensor term gives rise to two types of contributions . They are
proportional
to $f_{\rho}^2$ and $f_{\rho} g_{\rho}$ , and they are denoted , respectively ,
by $ (X^{(T)},Y^{(T)})$ and $(X^{(VT)} , Y^{(VT)})$ . Furthermore , they can be
splited into
timelike and spacelike component .

The time component of $X^{(T)}$ and $Y^{(T)}$ is written as
\begin{eqnarray*}
\left( \begin{array}{l}
 -X^{(T)}_0(r)  \\
  Y^{(T)}_0(r)
\end{array} \right)
& = & - (\frac{f_{\rho}}{2 M})^2 m_{\rho}^2 j_a^{-2} \sum_b (2-\delta_{q_a
q_b}) \{ \frac{j_a^2 j_b^2}{8 \pi}
\frac{(G_a F_b + F_a G_b)_r}{m_{\rho}^2 r^2}
\left( \begin{array}{l}
 -G_b(r) \\
  F_b(r)
\end{array} \right) \\
& - & m_{\rho} \sum_L \hat{L}^{-4} |<a||Y_L||b>|^2 \sum_{L_1,L_2}
\left( \begin{array}{l}
G_b(r)[\tilde{\kappa}_{ab}-\alpha(L_1)]  \\
F_b(r)[\tilde{\kappa}_{ab}+\alpha(L_1)]
\end{array} \right) i^{L_2-l_1} ~~~~(B.6)\\
&\times& \int_0^{\infty} [\{ \tilde{\kappa}_{ab} + \alpha(L_2) \} G_a F_b -
\{ \tilde{\kappa}_{ab} - \alpha(L_2) \} F_a G_b ]_{r'} R_{L_1,L_2}(m_\rho r ,
m_\rho r') dr' \} {\ }.
\end{eqnarray*}
where $\tilde{\kappa}_{ab} = \kappa_a - \kappa_b $ .

The space component of $X^{(T)}$ and $Y^{(T)}$ is
\begin{eqnarray*}
\left( \begin{array}{l}
 -X^{(T)}(r)  \\
  Y^{(T)}(r)
\end{array} \right)
& = & 6 (\frac{f_{\rho}}{2 M})^2 m_{\rho}^2 j_a^{-2} \sum_b (2-\delta_{q_a
q_b})
\left( \begin{array}{l}
 -F_b(r)  \\
  G_b(r)
\end{array} \right)
\sum_{L J L_1 L_2} f_{LJ}(L_1) f_{LJ}(L_2) \\
&\times& \left( \begin{array}{l}
 <a'||T_{J L_1}||b'> \\
 <a||T_{J L_1}||b>
\end{array} \right)
 \int_0^{\infty} [ <a||T_{J L_2}||b> G_a G_b -<a'||T_{J L_2}||b'> F_a F_b
]_{r'}  \\
&{\ }& [m_{\rho} R_{L_1 L_2}(m_\rho r , m_\rho r') - \frac{\delta
(r-r')}{m_\rho ^2 r'^2}] dr' ,~~~~~~~~~~~~~~~~~~~~~~~~~~~~~~~~~~~(B.7)
\end{eqnarray*}
where we have introduced
$$ f_{LJ}(L_1) = \hat{L} \hat{L}_1
\left( \begin{array}{lll}
L_1  &  L  &  1 \\
0    &  0  &  0
\end{array} \right)
\left\{ \begin{array}{lll}
L_1  &  L  &  1 \\
1    &  1  &  J
\end{array} \right\} $$

   The time components of VT contributions are
\begin{eqnarray*}
\left( \begin{array}{l}
-X^{(VT)}_0(r)  \\
 Y^{(VT)}_0(r)
\end{array} \right)
& = & (\frac{g_{\rho} f_{\rho}}{2M}) m_{\rho}^2 \hat{j}_a^{-2} \sum_b
(2 - \delta_{q_a q_b}) \sum_{L L_1} (-1)^{L_1} \hat{L}_1
\left( \begin{array}{lll}
L_1 & L & 1 \\
0   & 0 & 0
\end{array} \right) \\
& \{ &  \left( \begin{array}{l}
-G_b(r) <a'||T_{L L_1}||b>  \\
 F_b(r) <a||T_{L L_1}||b'>
\end{array} \right)  \\
&\times& \int_0^{\infty} [ <a||Y_L||b> G_a G_b + <a'||Y_L||b'> F_a F_b ]_{r'}
S_{L L_1} (r,r') dr'~~~~(B.8)  \\
& + & \left( \begin{array}{l}
-F_b(r) <a'||Y_{L}||b'>  \\
 G_b(r) <a||Y_{L}||b>
\end{array} \right)      \\
&\times& \int_0^{\infty} [ <a||T_{L L_1}||b'> G_a F_b + <a'||T_{L L_1}||b> F_a
G_b ]_{r'} S_{L L_1} (r',r) dr' \} .
\end{eqnarray*}

The space component of VT contribution are
\begin{eqnarray*}
\left( \begin{array}{l}
-X^{(VT)}(r)  \\
 Y^{(VT)}(r)
\end{array} \right)
& = & - \sqrt{6} (\frac{g_{\rho} f_{\rho}}{2M}) m_{\rho}^2 \hat{j}_a^{-2}
\sum_b
(2 - \delta_{q_a q_b}) \sum_{J L L_1} (-1)^{J} f_{LJ}(L_1) \\
& \{ & \left( \begin{array}{l}
 F_b(r) <a'||T_{J L_1}||b'> \\
 G_b(r) <a||T_{J L_1}||b>
\end{array} \right)  \\
&\times& \int_0^{\infty} [<a||T_{JL}||b'>G_a F_b - <a'||T_{JL}||b> F_a G_b
]_{r'} S_{L L_1}(r,r') dr'~(B.9) \\
& + & \left( \begin{array}{l}
 G_b(r) <a'||T_{J L}||b> \\
 F_b(r) <a||T_{J L}||b'>
\end{array} \right)  \\
&\times&\int_0^{\infty} [<a||T_{J L_1}||b>G_a G_b - <a'||T_{J L_1}||b'> F_a F_b
]_{r'} S_{L L_1}(r,r') dr' \}  .
\end{eqnarray*}
where
$$ S_{L L_1}(r , r') = \tilde{I}_{L_1}(m_\rho r) \tilde{K}_L(m_\rho r') \theta
(m_\rho (r'-r))
- \tilde{I}_{L}(m_\rho r') \tilde{K}_{L_1}(m_\rho r) \theta (m_\rho (r-r')) .
$$

\newpage
\begin{center}
{\bf{\Large{Figure captions}}}
\end{center}
\begin{description}
\item[Figure 1.] Scalar and vector potentials U$_{s}$ and U$_{0}$ as
functions of the density in nuclear matter. The circles are the DBHF
results using Bonn A potential[1]. The curves are obtained in terms
of interpolations and extrapolations.
\item[Figure 2.] Density dependent coupling constants of $\sigma$ and
$\omega$, g$_{\sigma}$ and g$_{\omega}$. They are deduced by
reproducing the scalar and vector potentials of the DBHF results at
each density from RMF analyses (a) and RHF analyses (b). The solid
and dotted curves in (b) are corresponding  to the cases of $\sigma$+
$\omega$ and $\sigma$+$\omega$+$\pi$+$\rho$ in RHF. The circles on
curves refer to the density points of the DBHF results.
\item[Figure 3.] Charge density distribution of various nuclei . The cureves
are the results of RDH (dashed one), RDHF with $\sigma$ + $\omega$
only (dash-dotted) and RDHF with $\sigma$ + $\omega$ + $\pi$ + $\rho$
( solid for 3C and dotted for 3A ) and RHF with $\sigma$ + $\omega$ + $\pi$ +
$\rho$ (dence-dotted
one ) .
\item[Figure 4.] Binding energy versus 1/$r_{ch}$ for $^{16}$O and $^{40}$Ca .
The
dash-dotted lines is taken from the work of K{\" u}mmel et al. \cite{kum78}.
The dotted and dashed lines are obtained by RHF3 with changing $\sigma$
meson mass for $K_F$ = 1.4 fm$^{-1}$ (1) and $K_F$ = 1.3fm$^{-1}$ (2) ,
respectively .
The solid lines are the results of RDHF3C . The experimental data are
displayed by a star .
\item[Figure 5.] Difference between charge densities of $^{40}$Ca and $^{48}$Ca
multiplied by $r^{2}$ .
The dashed curve corresponds to the RDH , the dotted curve to the RDHF1 .
The solid line corresponds to the RDHF3C .
The shaded area indicates the experimental data.
\item[Figure 6.] Same as Fig.4 for difference between neutron densities of
$^{48}$Ca and $^{40}$Ca.
\item[Figure 7.] $\Delta_{np}$ versus the asymmetry parameter for various
cases and the DBHF approach taken from ref.\cite{boe94}
\end{description}


\newpage
\begin{thebibliography}{10}
\bibitem{ser86}  B.D.Serot and J.D.Walecka,
\newblock{\it Adv. in Nucl. Phys.} 16, ed. J.W.Negele and
  E.Vogt (Plenum, New York, 1986).
\bibitem{gam90}  Y.K.Gambhir, P.Ring and Thimet
\newblock{\it Ann. Phys} 551(1990),129.
\bibitem{tok91}  H. Toki, Y.Sugahara, D. Hiratu and I.Tanitata
\newblock{\it Nucl. Phys} A524(1991),633.
\bibitem{sha94}  M.M.Sharma, G.A.Lalazissia, W.Hillebrandt and P.Ring
\newblock{\it Phys.Rev.Lett} 72(1994),1431.
\bibitem{bou87}  A.Bouyssy, J.-F.Mathiot, N.Van Giai and S.Marcos,
\newblock{\it Phys. Rev.}  C36, 380 (1987).
\bibitem{ber93}P. Bernardos et al.,
\newblock{\it Phys.Rev.} C48, 2665 (1993).
\bibitem{zha91}  J.K.Zhang, and D.S.Onley,
\newblock{\it Phys. Rev.} C43, 942 (1991).
\newblock{\it Phys. Rev.} C46, 1677 (1992).
\newblock{\it Phys. Rev.} C48, 2697 (1993).
\bibitem{coe70}   F.Coester, S.Cohen, B.D.Day and C.M.Vincent,
\newblock{\it Phys.Rev.} C1, 769 (1970).
\bibitem{kum78}  H.K{\" u}mmel,K.H.Luhrmann,and J.G.Zabolitzky,
\newblock{\it Phys. Rep} 36, 1(1978).
\bibitem{anu83} M.R.Anustusio, L.S.Celenza, W.S.Pong and C.M.Shakin,
\newblock{\it Phys.Rep.} 100, 327 (1983);
\bibitem{haa87}  B.ter Haar and R.Malfliet,
\newblock{\it Phys. Report} 149, 207 (1987).
\bibitem{hor87}  C.J.Horiwize and B.D. Serot,
\newblock{\it Nucl.Phys} A464, 613 (1987).
\bibitem{bro84} R.Brockmann and R.Machleidt,
\newblock{\it Phys.Lett.} B149, 283 (1984);
\newblock{\it Phys.Rev.} C42, 1965 (1990).
\bibitem{mut90} H.M{\" u}ther, R.Machleidt and R.Brockmann,
\newblock{\it Phys. Rev.} C42, 1981 (1990).
\bibitem{gmu91}   S.Gmuca,
\newblock{\it J of Phys.} G17, 1115 (1991).
\newblock{\it Z. Phys.} A342, 387 (1992).
\bibitem{bro92} R.Brockmann and H.Toki,
\newblock{\it Phys. Rev. Lett.} 68, 3408 (1992).
\bibitem{fri93} R.Fritz, H.M{\" u}ther, and R.Machleidt,
\newblock{\it Phys. Rev. Lett.} 71, 46 (1993).
\bibitem{ma94c}  Zhong-yu Ma, Hua-lin Shi and  Bao-qiu Chen,
\newblock{to be publiched in Chinese Phys. Lett. }
\bibitem{fri94} R.Fritz and H.M{\" u}ther,
\newblock{\it Phys. Rev. } C49, 633 (1994).
\bibitem{boe94} H. F. Boersma and R. Malfliet,
\newblock{\it Phys. Rev. } C49, 233 (1994); C49, 1495 (1994).
\bibitem{ma94p} Zhong-yu Ma, Hua-lin Shi and  Bao-qiu Chen,
\newblock{to be publiched in Phys. Rev. C50(1994) }
\bibitem{mac89} R. Machleidt,
\newblock{\it Adv. Nucl. Phys.} 19, 189 (1989)
\bibitem{hor83}   C. J. Horowitz and B. D. Sorot,
\newblock{\it Nucl. Phys.} A399, 529 (1983)
\bibitem{fet71}  A. L. Fetter and J. D. Walecka,
\newblock{\it Quantum Theory of Many-Particle System (McGram-Hill, New York,
1971) }
\bibitem{ma88}Zhong-yu Ma, Ping Zhu, Y. Q. Gu and Y. Z. Zhou,
\newblock{\it Nucl. Phys.} A490, 619 (1989).
\bibitem{ma94}  Zhong-yu Ma and Bao-qiu Chen,
\newblock{\it Commun. Theor. Phys.} 21, 59 (1994).
\bibitem{fri-em} R. Fritz ,
\newblock{private communication.}
\end{thebibliography}
\end{document}